\documentclass[pra,aps,twocolumn]{revtex4}

\usepackage{graphicx}
\usepackage{subfigure}

\begin{document}

\title{Demonstration of conditional quantum phase shift between ions in a solid}
 \author{J. J. Longdell}
 \email{jevon.longdell@anu.edu.au}
 \author{M. J. Sellars}
 \author{N. B. Manson}
 \affiliation{Laser Physics Centre, Research School of Physical
    Sciences and Engineering, Australian National University,
    Canberra, ACT 0200, Australia.}
\begin{abstract}
Due to their potential for long coherence times, dopant ions have long
been considered promising candidates for scalable solid state quantum
computing. However, the demonstration of two qubit operation  has proven to be problematic,
largely due to the difficulty of addressing closely spaced ions. Here
we use optically active ions and optical frequency addressing to demonstrate a
conditional phase shift between two qubits. 
\end{abstract}

\maketitle

Solid state quantum computing is a field of active research spurred in
part by the expectation that it will be easier to scale a solid state
system to a large number of qubits. To date the only experimental
demonstrations approaching that of a two qubit gate operation
\cite{yama03} have used qubits based on electron charge. While single
electron charges are measurable, their strong interaction with the
environment means that the coherence times are very short making
necessary extreme experimental techniques.

In this work we demonstrate a conditional quantum phase shift in a
two-qubit system, based on local interactions between optically active
centers in a solid.  The qubits were based on optical transitions of
the dopant ions.  The gate operations are performed with sequences of
precise optical pulses in a manner analogous to techniques used in
nuclear magnetic resonance. 

The material chosen for the demonstration was Eu doped yttrium
orthosilicate (YSO). This material was selected because of the narrow
homogeneous linewidth of the $^7$F$_0\leftrightarrow ^5$D$_0$
transition of the Eu$^{3+}$ ions at 579~nm, which can be as narrow as
100~Hz \cite{ultraslow}.  The long coherence time associated with this
narrow line width simplifies the implementation of the optical pulse
sequences as it allows the use of microsecond time scale pulses to
control the gate operations.  These pulses can be generated by simple
modulation of the output of a narrowband continuous wave laser.

The dominant interaction between the europium ions is a non-resonant
electric dipole-dipole interaction.  The Eu$^{3+}$ dopant ion
possesses a permanent electric dipole moment that depends on the
electronic state of the ion, therefore exciting one ion induces linear
Stark shift in the optical frequencies of surrounding ions.  For
europium ions in adjoining Y sites, separated by 0.5 nm, the
interaction is expected to be of the order of 10~GHz \cite{ohls02}.
The interaction falls off as the inverse of the cube of the separation
between the ions and therefore is larger than the ions' homogeneous
linewidth out to distances of approximately 500~nm.  A feature of the
electric dipole-dipole interaction which will assist in scaling to a
larger number of qubits, is that it can be effectively turned off by
storing the quantum information in the dopants' optically addressable
hyperfine states, this has been demonstrated experimentally
\cite{anabelsthesis}. Another benefit is that the hyperfine have long
coherence times, tens of seconds have recently been measured for these
hyperfine transitions \cite{elliot_personal}.

In a europium doped YSO crystal the europium ions substitute for the
yttrium ions in the lattice and are distributed randomly throughout
the available sites in the sample.  In general, each of these ions has
a unique optical transition frequency that is determined by the local
crystal environment (i.e. inhomogeneous broadening).  In principle it
would be possible to implement two qubit gate operation on a single
pair of closely spaced europium ions, addressing each ion by their
different optical frequencies.  In practice the optical detection of a
single europium ion will be difficult due to the weak oscillator
strength.  Therefore, to increase signal strength, a high
concentration of dopant ions is incorporated into the crystal, and
from this starting ensemble of ions we select out a sub-ensemble with
the desired characteristics.

The ability to select groups of ions that can be used as single qubits has
been demonstrated previously \cite{pryd00,tomog}. The general concept
is to use optical pumping to an auxiliary hyperfine level to select a
subgroup of ions with the desired characteristics. Here, a spectral
hole is first burned into the broad inhomogeneous optical linewidth of
the dopant ion transition.  The hole width is usually chosen to be in
the range of 1~MHz, so that it will be transparent to optical pulses
having a smaller bandwidth.  In the center of this broad spectral
hole, a narrow anti-hole is then created.  This is achieved using a
narrow band laser to selectively excite a transition from one of the
other ion ground-state sublevels to a convenient excited state,
thereby optically pumping some of the ions back. This gives a narrow
absorption feature (anti-hole) within the broad spectral hole.  The width of
this narrow anti-hole is determined by the inhomogeneous width of the
spin transition, which is on the order of 100~kHz.

In principle, all the ions in this narrow anti-hole will respond in
the same way to a sufficiently short laser pulse.  However, in
practice, non-uniformities in the laser intensity cause each ion to
respond at a different rate (Rabi frequency) to an applied laser
pulse.  To overcome this limitation, a series of what are nominally 2$\pi$ laser pulses
are applied, these leave ions in a particular part of the
laser beam in the ground state.  Any ion that does not respond with
the correct Rabi frequency is left partially excited and is optically
pumped to an auxiliary state after a few cycles.  When driven with sufficiently short
pulses (less than 10~$\mu$s), the remaining ions can be viewed as an
ensemble of identical single qubits.  For long periods of evolution,
$\pi$ pulses are needed to rephase the ensemble, because the remaining
100~kHz inhomogeneous broadening means that the optical coherence will
dephase over times of 1/(100~kHz). This rephased coherence, or optical echo, is also
used for quantum state readout because it produces a measurable
optical output beam whose properties are completely determined by the
quantum state of the ensemble.

\begin{figure}
  \includegraphics[width=0.45\textwidth]{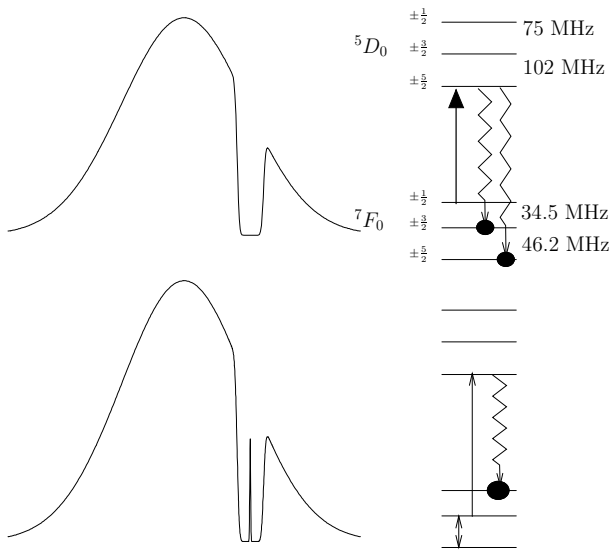}
\caption{Preparation sequence for distilling an ensemble of ions with
  well-defined properties out of an initially inhomogeneous
  distribution.  First a broad hole is burned via spectral hole
  burning.  (b) Next a narrow anti-hole is created in the center of
  the broad hole using optical pumping with a narrow band laser.
  Finally (not shown) ions in a particular part of the beam are
  selected using a series of 2$\pi$ pulses. The absorption profiles
  shown are not to scale. The inhomogeneous linewidth was about 5~GHz,
  the broad spectral hole was 1~MHz wide and the narrow anti-hole was
  100~kHz wide.
  \label{fig:figone}}
\end{figure}

Here we extend the single qubit selection process to the selection of a
group of ion pairs and then use this ensemble for a two-qubit
demonstration.  It should be noted that although we use ensembles, the
ability to place every atom in an identical initial state enables this
demonstration to be performed without the controversial pseudo-pure
states that are used in liquid NMR quantum computers.

For two-qubit operation, two of the anti-holes described above where
created, with one anti-hole used as the target qubit and the second
used as a control.  The interaction between the ions that is used here
was first characterised by Huang et al.\ \cite{huan89} and can be
observed using a technique known as echo demolition.  First, a
two pulse photon echo sequence is applied to the target ensemble
consisting of the usual $\pi$/2 pulse followed after a delay by the
rephasing $\pi$ pulse. For echo demolition, a perturbing pulse is
applied to the control ensemble at the same time as the rephasing
pulse is applied to the target.  The effect of the perturbing pulse is
to impart a frequency shift, and hence a phase shift, to each of the
ions in the target ensemble such that the rephased coherence ends up
phase shifted relative to the initial coherence. Since the
perturbation is only applied during the rephasing part of the echo
sequence, its effect is not canceled, as would be the case if it were
applied immediately after the initial pulse.

\begin{figure}
  \centering
   \subfigure[]{\label{fig:intrabia}\includegraphics[width=0.45\textwidth]{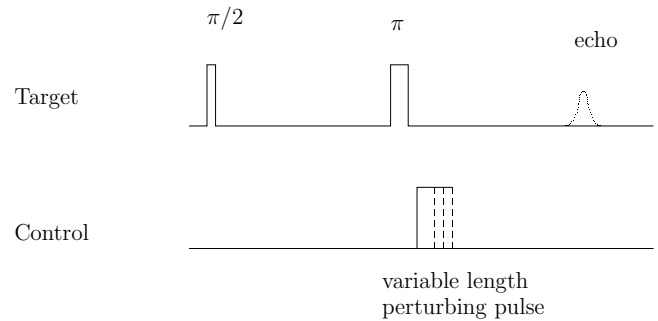}}
  \subfigure[]{\label{fig:intrabib}\includegraphics[width=0.45\textwidth]{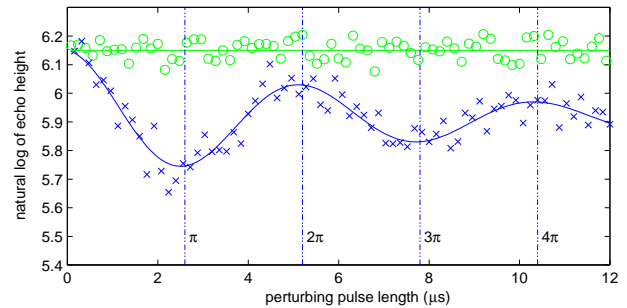}}
 \caption{(a) Pulse sequence for observation of control ion nutation
  on target ion echo. (b) Rabi flops of the control ion observed on
  the echo amplitude of the target ion ("x" symbols).  For comparison,
  the "circle" symbols correspond to the control ion not being
  excited.  To achieve such a clear Rabi nutation signal, the control
  ion was optically prepared by a sequence of 10 2$\pi$ pulses, as
  described above.} 
  \label{fig:intrabi}
\end{figure}

FIG.~\ref{fig:intrabi} shows the effect of the  interaction between
two single qubit ensembles. Excitation of the control ions shifts the
frequency of the target ions, and as this random frequency shift is
only present for half of the photon echo sequence it stops the target
ensemble rephasing to form an echo. In the figure Rabi flopping can be
seen in the target ensemble's echo as the length of the pulse applied
to the control ensemble is varied --- driving the control to the excited
state destroys the echo whereas cycling back to the ground state
leaves the echo unchanged.

\begin{figure}
  \centering
  \includegraphics[width=0.45\textwidth]{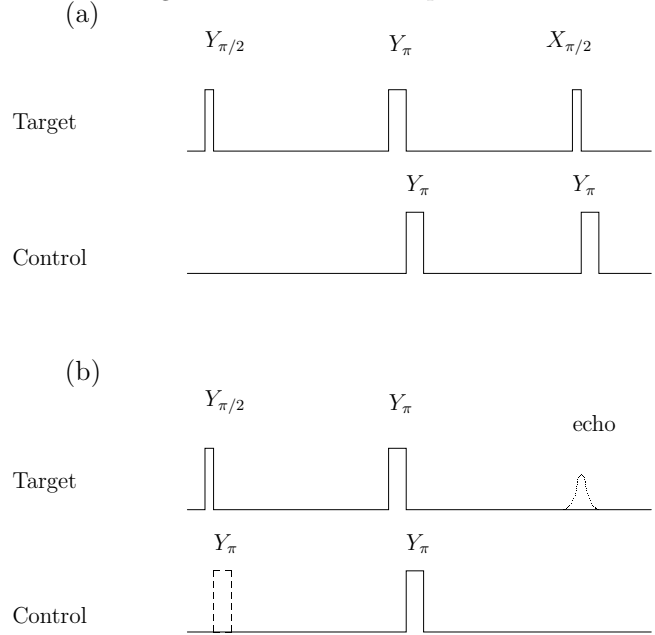}
  \caption{Pulse sequence used to (a) select target ions based on their interaction with the control and (b) demonstrate the conditional phase shift.
  \label{fig:pulseseq}}
\end{figure}

In order to perform two qubit logic operations, the echo demolition
must be turned into a controlled phase shift that does not decrease
the echo magnitude.  This is accomplished by using spectral hole
burning and the dipole coupling-induced frequency
shift to further prepare the target ensemble.  The target ions that do not
shift in frequency by the correct amount are optically pumped to a
passive level.  Conceptually this technique has been previously
proposed \cite{ohls02} using a CW approach. However it was
found that pulsed techniques borrowed from NMR spectroscopy were
needed to resolve the induced frequency shifts from the anti-hole's inhomogeneous
broadening.  Specifically, an echo pulse sequence is constructed that
returns a target ion, with the correct control ion interaction, back
to the ground state.  However, target ions with no control ion or with
a control ion coupling of the wrong strength are not returned
completely to their ground state, and are therefore optically pumped
away after several cycles.  The pulse sequence used for selecting the
ion pairs is illustrated in FIG.~\ref{fig:pulseseq}, the same pulse
sequence can be used for CNOT gate in NMR quantum computing. As the
control ions are not initially excited, the target ions with the
correct interaction strength will be returned to the ground state.

Because of the rephasing pulses the spectral resolution of this
technique is limited by the optical coherence time, which is orders of
magnitude narrower than the width of the anti-holes used. 
It should be noted that in this inital experiment the selection
process was only applied to the target ions.
As a result there are ions left in the control anti-hole that don't interact
with an ion in the target. 
A ensemble that is much closer to a true collection of pairs could be
achieved using the same pulse sequence but with the roles of the two
qubits reversed on alternate repetitions.

\begin{figure}
  \centering
    \includegraphics[width=0.45\textwidth]{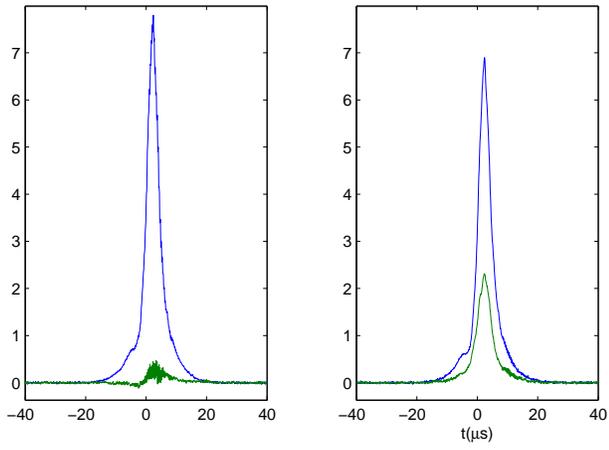}
  \caption{Conditional quantum phase shift  demonstration. The data
    for in-phase and in-quadrature echo signals are shown without (left)
    and with (right) the control ion initially excited.  Significantly
    the phase shift is produced without noticeable decay in the echo
    magnitude, given by the RMS sum of in-phase and in-quadrature
    signals. In other words the conditional phase shift is achieved
    without increaseing the affecting state purity.  The phase shift
    in this very first experiment is about 20-degrees, but in
    principle can be made larger than $\pi$ by upgrading the
    experiment.}
  \label{fig:phaseshift}
\end{figure}

The operation of the conditional quantum phase shift is shown in
FIG.~\ref{fig:phaseshift}, where the preparation pulse sequence
described in FIG.~\ref{fig:pulseseq} has been used.  The in-phase
and in-quadrature parts of the echo signal are plotted (superimposed) in
FIG.~\ref{fig:phaseshift}(a) as a function of time.  When the
control ion is excited, the echo is phase shifted.  Significantly, the
echo magnitude is essentially the same in both cases, thus, the echo
is no longer demolished by the perturbing pulse, but instead
experiences a conditional phase shift (i.e.  conditional quantum phase
gate).  The phase shift in this data is about 20 degrees, but in
principle can be made larger by further refinement of the state
preparation process.

Here we have started with a random ensemble of europium dopants in
yttrium orthosilicate, and we used spectral holeburning techniques to
select out ions from one ensemble that have a particular interaction
strength with ions in another. Using this prepared system we
demonstrated a phase shift in one qubit conditional on the state of
the other. The specific approach of selecting ensembles from randomly
distributed ions is not scalable because the requirements for the
ensemble get more and more stringent as the number of qubits increase.
To achieve scalable quantum computing it will be necessary to either
implement single site detection or develop a method of creating
identical clusters of optically active ions. Single site detection
will be difficult for rare-earth ions due to their weak oscillator
strength, however the ion-ion interaction used here will be present in
any solid state impurity site with a permanent electric dipole moment.
This includes the NV-centre in diamond where single site detection is
well advanced \cite{grub97}.

The authors wish to thank Dr.~P.~R.\ Hemmer for his helpful comments and
acknowledge financial assistance from the Australian Defense Science
and Technology Organisation, the Australian Research Council and the
US Defense Advanced Research Projects Agency.


\end{document}